# DevOps and General Developers: Insights from Stack Overflow's 2023 Survey

Hasan Hameed Hasan Ahmed Abdulla
College of Information Technology, University of Bahrain, Sakhair, Kingdom of Bahrain

Fatema Hasan AlJazeeri
College of Information Technology, University of Bahrain, Sakhair, Kingdom of Bahrain

Fawzi Abdulaziz AlBalooshi
Department of Computer Science, College of Information Technology, University of Bahrain, Sakhair, Kingdom of Bahrain

Jaflah Hassan Al-Ammary
Department of Information Systems, College of Information Technology, University of Bahrain, Sakhair, Kingdom of Bahrain

## Abstract

**Purpose** – To investigate the distinct roles of DevOps specialists and general software developers, examining their varying use of tools, technologies, methodologies, and demographics in the current software development environment. In addition, to differentiate these two professional groups regarding their unique contributions and challenges in the field.
**Design/Methodology/Approach** – The research uses a quantitative approach to analyze data from the Stack Overflow 2023 Developer Survey. It focuses on a comparative analysis of technological preferences, demographic information, and professional experiences between DevOps specialists and general developers, highlighting key trends and differences. The data analysis was conducted using Python's Pandas library for data analysis.
**Findings** – The research indicates no significant difference in the tool and technology preferences between DevOps specialists and general software developers, highlighting their complementary roles. DevOps specialists and general software developers use tools like Docker and Kubernetes, emphasizing efficiency and automation. While general developers employ diverse tools for various role demands, demographic trends show younger general developers and mid-career DevOps professionals. This age range reflects growing experience in DevOps, and both groups are adapting to remote and hybrid work models in the evolving tech industry.
**Practical Implications** – This research offers perspectives on the dynamic roles within software development, emphasizing the growing importance of DevOps. It is a valuable resource for academic and industry professionals to understand the evolving dynamics in software development roles.
**Originality/Value** – This research fills a significant gap in the existing literature. It serves as a valuable resource for academic and industry professionals, aiding in comprehending the evolving dynamics of software development roles and the increasing significance of DevOps in the industry.

**Keywords:** DevOps, Software Development, Stack Overflow Survey, Technology Trends

# 1. Introduction

The evolution of software development methodologies over recent decades is well-documented, transitioning from structured approaches like the Waterfall model to more dynamic and integrated methods. Initially, software development was characterized by rigid, linear methodologies such as the Waterfall model, which adhered to a sequential design process (Vollmar & Bohm, 2022). These methods, while structured, faced criticism for their inflexibility and inability to adapt to changing requirements.

In the early 2000s, the software development industry underwent a significant shift with the adoption of Agile methodologies. Agile methodologies emphasize adaptability, customer-centric development, and iterative processes, departing from traditional software development approaches (Marrero & Astudillo, 2021). Agile enhanced collaboration among teams and stakeholders, contributing to a more integrated and responsive software development environment.

The emergence of DevOps further advanced this evolution. DevOps, a term coined in 2009, extended beyond a mere methodology, embodying a cultural and professional ethos emphasizing collaboration and integration in software development (Rodygina et al., 2023). This approach focuses on continuous communication, shared responsibilities, and aligning development with business objectives and customer satisfaction, which are critical elements of DevOps (Thantharate, 2023).

DevOps integrates various functions into a cohesive workflow, promoting continuous integration, rapid deployment, and process automation, reflecting a shift towards a more agile, user-focused development ethos (González et al., 2023). It necessitates a cultural shift among stakeholders in software development and deployment. The widespread adoption of DevOps has significantly impacted the roles and relationships within organizations. It has evolved from a niche concept to a mainstream practice, integral to the operations of many organizations, highlighting a transition towards more efficient and streamlined development and operational processes (Hamzane & Badr, 2021). This mainstream integration underscores the importance of collaboration and continuous improvement in modern software development.

In the rapidly evolving field of software development, understanding professionals' distinct roles and contributions is essential. The DevOps movement embodies a novel paradigm in software development, deployment, and maintenance, distinguishing itself from traditional software development practices. This distinction is highlighted in the study by Ebert and Hochstein (2023), which discusses DevOps in practice, suggesting that DevOps specialists often possess unique competencies, tools, and methodologies (Ebert & Hochstein, 2023). Their role frequently transcends conventional development tasks, encompassing system administration, network operations, and continuous deployment. This multifaceted nature of DevOps roles is further elucidated by Gallagher and Lennon (2022), who explored architecting multi-cloud applications for high availability using DevOps, thereby underscoring the comprehensive skill set required in this field (Gallagher & Lennon, 2022). Additionally, the integration of agile methodologies within the DevOps framework, as discussed by Cardoso et al. (2021), serves to enhance both the technical and non-technical skills of practitioners, thereby reinforcing the unique position of DevOps in the software development landscape (Cardoso et al., 2021).

The comparison provides a multitude of valuable insights. Delineating the skillset and methodology sheds light on the specialized skills. Its approaches are essential for DevOps practitioners, particularly critical for organizations evolving with contemporary software development practices. Furthermore, the analysis has significant implications for education and training, assisting educators and trainers in designing programs that effectively prepare the upcoming cadre of software professionals. Additionally, an in-depth look at demographic factors such as age, employment status, and educational background offers a clearer understanding of the diversity within the software development community. This understanding is vital to creating inclusive, innovative, and efficient work environments. The demographic analysis also uncovers trends in professional development and career paths, providing invaluable information for industry leaders and policymakers invested in workforce development and strategic planning.

This study aims to uncover different facets of the software development industry, focusing on the distinctions between DevOps specialists and general developers. The comparison will focus on the differences in demographic factors and tool usage between the two categories.

The paper will be organized as follows: Section 2 will provide a literature review for the research area. Section 3 will discuss the paper's methodology. Sections 4 and 5 will present the results and discussion, respectively. Finally, the conclusion will be covered in section 6.

## 2. Literature Review

Integrating DevOps into software development has revolutionized the industry by fostering collaboration between development and operations teams (Cardoso et al., 2021). The DevOps approach aims to shorten the development lifecycle while ensuring that software updates and fixes align with business objectives, a sentiment echoed by Ebert and Hochstein (2023) in their discussion of DevOps implementation.

The adoption of tools and technologies is critical in the DevOps landscape. Beetz and Harrer (2022) highlight the evolution of GitOps as an iteration of DevOps, suggesting the constant evolution of tools and methodologies. Similarly, the work of Curty and Murta (2023) on the assignment of commits to releases underscores the technical intricacies involved in DevOps practices.

Grasping DevOps practitioners' demographic and professional characteristics is pivotal for discerning the dynamics within this domain. Research such as the study conducted by Weeraddana et al. (2023) compares diversity within the DevOps sector, illuminating the various backgrounds and experiences of professionals in this field. Additionally, the Stack Overflow Developer Survey, conducted annually, provides a comprehensive overview of the tools, technologies, and demographic profiles of developers globally, encompassing those in DevOps positions. This survey is crucial in detecting trends and transformations in the software development landscape by collecting pertinent data.

### *2.1 Evolution and Definition of DevOps*

The origins of DevOps trace back to the early 2000s, coinciding with the Agile movement. Agile methodologies highlighted the gap between development teams and IT operations, thus paving the way for a more integrated approach to software delivery (Cardoso et al., 2021). DevOps, coined by Patrick Debois in 2009, emerged as a response to the conflicting objectives of development and operations teams. It aimed to harmonize these goals by promoting collaboration between these traditionally separate entities (Gokarna & Singh, 2021). The concept gained momentum through forums and conferences like the inaugural DevOpsDays 2009. This event spread DevOps principles alongside the #DevOps hashtag (DevOps phenomenon, 2020).

As a methodology, DevOps significantly emphasizes culture, automation, lean management, and sharing. Ebert and Hochstein (2023) elucidate that DevOps integrates software development and IT operations, fostering enhanced communication and collaboration, which are fundamental to its success (Ebert & Hochstein, 2023). Incorporating process automation is a critical aspect of DevOps, as highlighted in the comprehensive review by Batra and Jatain (2020), which discusses the measurement-based performance evaluation of DevOps (Batra & Jatain, 2020).

Implementing technologies such as Continuous Integration/Continuous Deployment (CI/CD) pipelines, microservices, and infrastructure as code tools exemplify DevOps principles. These technologies enable automated testing, incremental deployments, and programmable infrastructure. Fox (2020) addresses the transformational impact of adopting DevOps, noting improvements in deployment frequency, recovery times, and the capacity for experimentation within organizations (Fox, 2020). The alignment of these advancements with the findings of Llisar and Oviedo (2021) is notable; they highlight the role of software reliability in a DevOps continuous integration environment (Llisar & Oviedo, 2021).

DevOps contrasts with traditional, plan-driven lifecycles by emphasizing rapid feedback and continuous evolution instead of linear phases and upfront planning. Such an approach leads to notable differences in speed, efficiency, and organizational culture in software development processes (Batra & Jatain, 2020). DevOps integrates tools, practices, and cultural philosophies to address software delivery challenges. This background provides context for further exploration of DevOps practices, tools, challenges, and demographics.

*2.2 Tools and Technologies in Software Development*

The evolution of software development tools, especially within DevOps, has significantly transformed software engineering. This transformation, characterized by integrating and automating various aspects of development and operations, aligns well with the DevOps philosophy (Ebert & Hochstein, 2023). Early software development utilized basic text editors and command-line interfaces for compiling. However, the 1970s saw a pivotal change with Integrated Development Environments (IDEs) like Turbo Pascal consolidating editing, building, debugging, and testing functions, streamlining development (Ebert & Hochstein, 2023).

As software development complexity has increased, there has been a corresponding escalation in the demand for tools capable of managing intricate workflows. Version control systems like Git have become indispensable for efficient tracking and managing collaborative code changes. This trend is underscored by Beetz and Harrer (2022) in their exploration of GitOps as an evolution of DevOps practices. The increased reliance on version control systems is aligned with the findings of Batra and Jatain (2020), who discuss the importance of testing frameworks, such as Selenium, in providing quick feedback and expediting cycles in DevOps methodologies. The emergence of cloud computing, containers, and API-driven toolchains has further propelled technological advancement, as Ebert and Hochstein (2023) noted in their examination of DevOps in practice. Tools like Terraform and Ansible have enabled programmatic cloud and server provisioning/management, reflecting a shift towards more agile and scalable deployment methodologies. Kubernetes, for example, has gained prominence for efficiently deploying containerized microservices. Additionally, CI/CD platforms like Jenkins have become critical in automating workflows, further facilitating this transformation in software development. Monitoring tools like Datadog have been essential in providing performance insights, underscoring the importance of continuous monitoring and improvement in DevOps environments, a point highlighted by Thompson and Erdogan (2020).

Tool integration plays a critical role in the efficacy of DevOps practices, with the selection of tools being pivotal in meeting organizational needs and fostering an environment conducive to automation, collaboration, and continuous practices. As indicated in the research, the integration of extensive toolchains significantly enhances operational efficiency, facilitates error detection, and optimizes cycle times. Popenţiu-Vlădicescu and Albeanu (2022) emphasize the value of automation in minimizing errors, accelerating deployment processes, and enabling timely feedback, which are crucial aspects of a robust DevOps strategy. Furthermore, integrating these tools improves visibility across the delivery pipeline, promoting better coordination and management of development and operations. This holistic approach to tool integration underscores its transformative impact on enhancing DevOps methodologies.

Adoption challenges like complexity, learning curves, and investment must be balanced with standardization and flexibility (Fox, 2020). Incorporating AI/ML is expected to enhance automation and intelligence in delivery further, requiring seamless integration into DevOps workflows (Rodygina et al., 2023). Tools have evolved from compilers to interconnected toolchains, becoming integral to modern practices like DevOps.

### *2.3 Comparative Studies between DevOps Specialists and General Developers*

In the emerging field of DevOps, the role of DevOps engineer has been clearly distinguished from that of traditional software developers (Curty & Murta, 2023; Ebert & Hochstein, 2023). DevOps engineers are primarily focused on infrastructure and pipeline development and maintenance, which are key components in the delivery of applications (Curty & Murta, 2023; Ebert & Hochstein, 2023). Their specialization often includes automation and system administration, with proficiency in tools such as Jenkins, Docker, and Kubernetes. This contrasts with the role of developers who are generally more involved in building applications using various programming languages, frameworks, and testing tools. Curty and Murta (2023) and Ebert and Hochstein (2023) both highlight these distinctions, emphasizing the unique responsibilities and skill sets required in the DevOps domain compared to traditional software development.

In the evolving software development landscape, developers, and DevOps engineers' distinct yet complementary roles have become increasingly apparent. Developers primarily focus on creating and developing applications. This assertion is supported by the work of Lee and Hong (2022), who emphasize the role of developers in urban development projects, highlighting their focus on application creation. Similarly, Oliveira et al. (2022) elaborate on improving developer productivity, particularly on the Internet of Things, reaffirming the developer's role in application development.

Toolchains vary significantly. DevOps engineers frequently use Infrastructure as Code tools, while developers more often use IDEs, version control, and code-sharing tools, underscoring each role's specialized nature (Popențiu-Vlădicescu & Albeanu, 2022). Workflows also differ. Developers typically follow Agile sprints or Waterfall, focusing on new features and bug fixes. DevOps engineers maintain deployable code repositories, implement CI/CD pipelines, and oversee monitoring (Gokarna & Singh, 2021).

Integrating DevOps improves lead time, deployment frequency, and mean time to recovery through increased automation and collaboration, enabling faster experimentation/recovery (Jamal, 2022). However, differences in priorities and lack of cross-training can create friction. Organizations implementing DevOps emphasize empathy and shared purpose across disciplines (Lavirotte et al., 2020).

Empirical research comparing DevOps specialists and developers reveals distinct but complementary roles. Studies by Gonzalez et al. (2023) and Jamal (2022) emphasize the unique skill sets required in DevOps, such as expertise in automation, system administration, and cloud technologies. Comparative studies highlight the varying workflows and tool preferences between these roles. DevOps engineers often use CI/CD tools and Infrastructure as Code, whereas developers predominantly work with IDEs, version control, and development frameworks. This distinction underlines the specialized nature of each role within the software development lifecycle.

As DevOps evolves, the roles become more intertwined as developers gain operational skills and DevOps engineers become more involved in codebases, reflecting a trend toward flexible, integrated teams (Grande & Vizcaíno, 2023). Comparative studies illuminate DevOps specialists' and developers' unique yet collaborative roles, setting the stage for examining DevOps demographics.

### *2.4 The Importance of Demographics in Software Development Roles*

The demographics of software professionals, including DevOps specialists and developers, present a diverse landscape shaped by factors like age, gender, education, geography, and work modes. A considerable proportion, around 67%, are under 40, reflecting the industry's youthful dynamism. This blend of youthful technology adoption and seasoned expertise fosters creativity and problem-solving within teams (Weeraddana et al., 2023).

Educational backgrounds vary. DevOps engineers have systems administration or computer science degrees, aligning with their infrastructure focus. Developers often specialize in software engineering and programming. Many self-taught professionals in both areas highlight the varied career pathways (Paez & De, 2022).

Geographical variations influence practices and tool preferences. For example, Chinese developers prefer proprietary tools to their Western counterparts' open-source inclinations. European developers adopt more formal processes than the agile models expected in the U.S., reflecting cultural differences (Gokarna & Singh, 2021).

Demographic research in software development provides insights into workforce diversity and its impact on the industry. Studies by Weeraddana et al. (2023) and Paez & De (2022) have explored the implications of age, gender, educational background, and geographic location on software development roles. These studies highlight the importance of diversity in fostering innovation and addressing the industry's evolving needs. For example, gender diversity has been linked to higher team productivity and collective intelligence, while geographical variations influence tool preferences and work methodologies.

The shift toward remote, contract, and freelance work modes significantly reshapes the sector by increasing flexibility but also posing communication and cohesion challenges. Proactive collaboration strategies are needed (Hubbs & Myren, 2023). Diversity in age, gender, education, geography, and work modes fosters a rich environment for innovation. Next, key themes and relationships will synthesize insights on the personnel shaping software development.

## 3. Methodology

This research paper aims to delineate the distinctions between DevOps specialists and general developers concerning their preferences for environments, tools, and technologies. Furthermore, it aims to scrutinize the impact of demographic factors on selecting DevOps specialists for tools and career paths within software engineering communities. In pursuit of these objectives, the paper employs a descriptive and quantitative approach, leveraging data extracted from the Stack Overflow Developer Survey 2023. This annual survey, renowned for its comprehensive scope, seeks to unveil the evolving trends within the developer community. The quantitative methodology of this study encompasses statistical techniques to succinctly summarize and analyze demographic data, educational backgrounds, and technology preferences.

The 2023 edition of the Stack Overflow Developer Survey was meticulously crafted to cater to a diverse range of participants within the realm of software development. This comprehensive survey targeted professional developers and extended its reach to encompass hobbyists and students actively engaged in the field. Designed to be as inclusive as possible, it aimed to capture a broad spectrum of experiences, skills, and perspectives. The participant recruitment for the study was conducted through various channels operated by Stack Overflow, targeting its extensive user base. The sampling strategy adopted was non-probabilistic, primarily relying on voluntary participation from individuals associated with the Stack Overflow community. This approach allowed a wide range of respondents to participate in the survey. (Stack Overflow Insights - Developer Hiring, Marketing, and User Research, n.d.)

The Stack Overflow Developer Survey was meticulously designed to encompass various topics relevant to the software development industry. It was divided into multiple sections, each focusing on a different aspect of the developer experience. Key areas covered in the survey included demographic information, such as age, geographic location, educational background, and professional qualifications. A significant portion of the survey was dedicated to understanding the technology usage patterns of the respondents, delving into their preferences for programming languages, frameworks, tools, and platforms. The survey also gathered data on the participants' work environment, exploring aspects such as employment status, work setting (remote, in-office, or hybrid), and job satisfaction. Additionally, the questionnaire delved into contemporary topics, such as AI sentiment, by gauging developers' attitudes and perspectives toward the increasingly significant role of artificial intelligence in software development.

The survey incorporated a mix of closed and open-ended questions to capture a comprehensive view of the respondents' experiences and opinions. This approach allowed for both quantitative and qualitative analysis of the data. The closed-ended questions provided structured data, enabling statistical analysis of trends and patterns. In contrast, the open-ended questions offered respondents the freedom to express their thoughts and experiences in more depth, providing richer, qualitative insights into the nuances of the software development industry. This combination of question types ensured a well-rounded understanding of the diverse perspectives within the developer community.

The Stack Overflow Developer Survey 2023 data analysis was executed using advanced statistical software, including Python's Pandas library. Python's Pandas library for data analysis was a strategic choice that significantly contributed to efficiently and effectively handling the Stack Overflow Developer Survey dataset. Its comprehensive functionalities facilitated a thorough exploration of the survey data, enabling the researchers to derive meaningful and actionable insights from the study.

## 4. Results

The survey was conducted online and accessible during May 2023, a strategic choice to ensure maximum participation. The online format of the survey was pivotal in its ability to reach a global audience, breaking geographical barriers and allowing developers from various regions and backgrounds to contribute their insights. This digital approach also facilitated ease of participation, encouraging a higher response rate.

The response to the 2023 survey was remarkable, with 89,184 individuals contributing their perspectives. This substantial participation rate underscores the survey's significance and relevance within the software development community. The large number of responses indicates the survey's broad reach and enhances the robustness and reliability of the collected data. With such a comprehensive dataset, the survey offers valuable insights into the trends, challenges, and evolving dynamics of the software development landscape as perceived by a broad and varied group of developers.

Before the analysis, the Stack Overflow Developer Survey dataset underwent a rigorous cleaning to ensure the highest data integrity and reliability. This step was crucial in preparing the data for accurate and meaningful analysis. The cleaning process focused on several key aspects.

The cleaning process involved addressing incomplete responses, handling outliers, and correcting inconsistencies. Incomplete responses were either rectified or excluded based on the extent of missing data. Outliers were scrutinized to distinguish genuine responses from anomalies, retaining the former to preserve data diversity. Inconsistencies and discrepancies were corrected to ensure logical consistency and standardized data formats.

These measures of data cleaning were instrumental in ensuring that the subsequent analyses were based on reliable and high-quality data. The meticulous data-cleaning process not only upheld the ethical standards of research but also contributed significantly to the credibility and trustworthiness of the study's findings.

The demographic analysis revealed insightful distinctions and parallels between DevOps Specialists and General Software Developers. Three key aspects were considered: age distribution, geographical distribution by country, and education levels. The findings are encapsulated in Tables 1, 2, 3, and 4, each providing a comprehensive view of these aspects.

**Table 1**: Age Distribution of DevOps Specialists and General Software Developers

| | **DevOps Specialists** | | **General Developers** | |
|---|---|---|---|---|
| **Age Group** | **Count** | **%** | **Count** | **%** |
| 18-24 years old | 204 | 14.72% | 8082 | 14.74% |
| 25-34 years old | 569 | 41.03% | 24069 | 43.91% |
| 35-44 years old | 420 | 30.28% | 14552 | 26.54% |
| 45-54 years old | 149 | 10.74% | 4983 | 9.09% |
| 55-64 years old | 39 | 2.81% | 1504 | 2.74% |
| 65 years or older | 3 | 0.22% | 197 | 0.36% |

**Table 2:** Statistical analysis of the age groups for DevOps Specialists and General Developers

| | **Count** | **Mean** | **Std. Deviation** | **Minimum** | **Maximum** |
|---|---|---|---|---|---|
| DevOps | 1387 | 34.36 | 9.41 | 21 | 21 |
| General Developers | 75485 | 34.45 | 10.23 | 65 | 65 |

Table 1 and Table 2 present the age breakdown of both groups. It highlights the age groups ranging from 18 to 65 years and older, comparing the count and percentage distribution within each group. The data reveals the concentration of developers in the 25-34 and 35-44 age ranges for both categories. For DevOps specialists, (71.31%) of the developers fall in the age ranges 25-34 and 35-44. At the same time, the general developers (70.45%) of the developers fall in the same range. In addition, the means for the two categories, as shown in Table 2, are similar. Therefore, the two categories have no remarkable difference in age factor.

**Table 3:** Geographical Distribution of DevOps Specialists and General Software Developers

| | **DevOps Specialists** | | **General Developers** | |
|---|---|---|---|---|
| **Country** | **Count** | **%** | **Count** | **%** |
| United States of America | 280 | 20.19% | 11217 | 20.45% |
| India | 137 | 9.88% | 8764 | 15.98% |
| Germany | 80 | 5.77% | 3664 | 6.68% |
| United Kingdom | 74 | 5.34% | 3461 | 6.31% |
| Canada | 53 | 3.82% | 2098 | 3.83% |

Table 3 findings reveal that the United States has the highest number of DevOps Specialists at 280 and General Developers at 11,217, constituting around 20% of their respective groups. India follows with a notable tilt towards General Software Development, having 137 DevOps Specialists and 8,764 General Developers. Germany's figures stand at 80 for DevOps Specialists and 3,664 for General Developers, while the United Kingdom records 74 DevOps Specialists and 3,461 General Developers, demonstrating a proportional engagement in tech roles. While presenting lower numbers, Canada mirrors this balanced distribution with 53 DevOps Specialists and 2,098 General Developers. The research underscores a consistent and balanced global representation of DevOps and General Software Development roles across the surveyed nations.

**Table 4**: Education Level Distribution of DevOps Specialists and General Software Developers

| | DevOps Specialists | | General Developers | |
|---|---|---|---|---|
| **Education Level** | **Count** | **%** | **Count** | **%** |
| Bachelor’s degree | 634 | 45.71% | 26485 | 48.25% |
| Master’s degree | 437 | 31.51% | 13195 | 24.04% |
| Some college/university study | 122 | 8.79% | 7153 | 13.03% |
| Associate degree | 49 | 3.53% | 1957 | 3.57% |
| Secondary school | 48 | 3.46% | 3421 | 6.23% |

The analysis of educational background in Table 4 shows that most of both DevOps Specialists and General Software Developers hold bachelor's degrees (45.7% and 48.2%, respectively). However, DevOps Specialists are more likely to have advanced degrees, with 31.5% holding master's degrees compared to only 24.0% of General Developers. The data also indicates that DevOps Specialists are slightly less likely to have just secondary education (3.5%) versus General Developers (6.2%). The findings suggest a strong preference for higher education among technology professionals, with some variance between roles. There is still a range of educational pathways represented in the field.

The analysis included an examination of professional attributes such as years of professional coding experience, employment status, and preferences for remote work. These attributes offer a deeper understanding of the two groups' professional environment and career progression. The professional attributes are concisely presented in Tables 5, 6, and 7.

**Table 5**: Professional Coding Experience of DevOps Specialists and General Software Developers

| | DevOps Specialists | | General Developers | |
|---|---|---|---|---|
| **Experience Range** | **Count** | **%** | **Count** | **%** |
| 0-1 years | 32 | 2.31% | 1310 | 2.39% |
| 0-5 years | 387 | 27.90% | 15628 | 28.47% |
| 6-10 years | 333 | 24.01% | 13140 | 23.94% |
| 11-20 years | 331 | 23.86% | 12016 | 21.89% |
| 21+ years | 159 | 11.46% | 6672 | 12.16% |
| 50+ years | 0 | 0.00% | 33 | 0.06% |
| Unknown | 145 | 10.45% | 6088 | 11.09% |

The analysis of professional coding experience, as presented in Table 5, reveals that a significant proportion of both DevOps Specialists and General Software Developers are in the early to mid-stages of their careers, with 27.9% and 28.5% having 0-5 years of experience, and 24.0% and 23.9% possessing 6-10 years, respectively. Additionally, there is a notable presence of seasoned professionals with 11-20 years of coding experience, accounting for 23.9% of DevOps Specialists and 21.9% of General Developers. This data underscores the diverse expertise levels among technology professionals. Moreover, there is no significant disparity in the years of experience between the two groups.

**Table 6**: Employment Status of DevOps Specialists and General Software Developers

| | **DevOps Specialists** | | **General Developers** | |
|---|---|---|---|---|
| **Employment Category** | **Count** | **%** | **Count** | **%** |
| Freelance | 105 | 7.57% | 6194 | 11.29% |
| Full-time | 1196 | 86.23% | 44160 | 80.46% |
| Part-time | 61 | 4.40% | 2706 | 4.93% |
| Seeking employment | 25 | 1.80% | 1803 | 3.28% |
| Not employed | 0 | 0.00% | 24 | 0.04% |

Employment status in Table 6 shows 86.2% of DevOps Specialists work full-time compared to 80.5% of General Software Developers, indicating a strong preference for full-time roles among DevOps Specialists. However, only 7.6% of DevOps Specialists work freelance versus 11.3% of General Developers, suggesting General Developers are more likely to take on freelance opportunities. The distribution of part-time roles is similar between the groups at around 5%. A slightly higher percentage of General Developers (3.3%) are seeking employment compared to DevOps Specialists (1.8%), potentially reflecting more dynamic career transitions among General Developers. Overall, the data reveals that most professionals in both categories work full-time, but General Developers show a more diverse pattern of employment arrangements, including higher freelancing.

**Table 7:** Remote Work Preferences of DevOps Specialists and General Software Developers

| | **DevOps Specialists** | | **General Developers** | |
|---|---|---|---|---|
| **Remote Work Preference** | **Count** | **%** | **Count** | **%** |
| Hybrid (some remote, some in-person) | 646 | 47.43% | 21607 | 40.73% |
| In-person | 123 | 9.03% | 7921 | 14.93% |
| Remote | 593 | 43.54% | 23523 | 44.34% |

The remote work preferences in Table 7 shows that 47.4% of DevOps Specialists prefer a hybrid model compared to 40.7% of General Software Developers, indicating a stronger preference for a mix of remote and in-person work among DevOps Specialists. In contrast, only 9.0% of DevOps Specialists want entirely in-person work versus 14.9% of General Developers, suggesting General Developers have a greater inclination toward in-person work. However, both groups have a high and nearly equal preference for fully remote work (43.5% of DevOps Specialists and 44.3% of General Developers). Overall, the data reveals a notable trend towards hybrid and fully remote work among technology professionals, with some variance between the two groups. Flexible and remote work options appear essential for the tech sector.

**Table 8**: Top 5 Technical Skills and Interests

| Category | Skill/Interest | DevOps Specialists Count | DevOps Specialists % | General Developers Count | General Developers % |
|---|---|---|---|---|---|
| Programming Languages | JavaScript | 453 | 6.27% | 18197 | 6.51% |
| | Python | 428 | 5.93% | 15520 | 5.54% |
| | SQL | 418 | 5.79% | 13738 | 4.91% |
| | Bash/Shell/PowerShell | 381 | 5.28% | 8945 | 3.19% |
| | TypeScript | 343 | 4.75% | 6727 | 2.40% |
| Databases | MySQL | 265 | 3.67% | 12324 | 4.41% |
| | PostgreSQL | 249 | 3.45% | 9791 | 3.50% |
| | MongoDB | 229 | 3.17% | 6860 | 2.45% |
| | Redis | 199 | 2.76% | 4533 | 1.62% |
| | SQLite | 192 | 2.66% | 8342 | 2.98% |
| Platforms | AWS | 821 | 24.86% | 23065 | 24.06% |
| | Docker | 588 | 17.81% | 16789 | 17.54% |
| | Linux | 488 | 14.79% | 15194 | 15.89% |
| | Windows | 303 | 9.19% | 10312 | 10.76% |
| | Kubernetes | 295 | 8.94% | 6825 | 7.15% |

Table 8 findings reveal significant similarities in technical proficiencies between DevOps Specialists and General Software Developers. In terms of programming languages, JavaScript is the most common skill for both groups, with 453 DevOps Specialists (6.27%) and 18,197 General Developers (6.51%) proficient in it, followed closely by Python, SQL, and Bash/Shell/PowerShell. In the Databases category, MySQL leads among both groups, used by 265 DevOps Specialists (3.67%) and 12,324 General Developers (4.41%). In the Platforms category, AWS is the most used by both DevOps Specialists (821, 24.86%) and General Developers (23,065, 24.06%), followed by Docker, Linux, Windows, and Kubernetes, all of which show significant usage patterns in both groups. These findings emphasize the critical technical skills required in the current tech industry, particularly focusing on cloud platforms, containerization, and programming languages.

**Table 9:** Comparative Analysis of Yearly Compensation Between DevOps and General Software Developers

| Type | Yearly Compensation Mean | Yearly Compensation Median |
|---|---|---|
| DevOps Specialists | $104,563 | $79,325 |
| General Software Developers | $103,110 | $74,963 |

In Table 9, the comparison of yearly compensation for DevOps Specialists and General Software Developers highlights that, on average, DevOps Specialists earn more than their counterparts in general software development. The mean yearly compensation for DevOps Specialists is $104,563, slightly higher than the average of $103,110 for General Software Developers. Additionally, the median yearly compensation for DevOps Specialists stands at $79,325, compared to $74,963 for General Software Developers. These findings indicate that specialized skills in DevOps may lead to higher financial rewards, emphasizing the increased value placed on the specialized skills and expertise of DevOps roles in terms of financial compensation.

An analysis of the top 10 tools used by DevOps Specialists and General Software Developers is presented in Table 10.

**Table 10:** Top 10 Tools used by DevOps Specialists and General Software Developers

| | **DevOps Specialists** | | **General Developers** | |
|---|---|---|---|---|
| **Tool** | **Count** | **%** | **Count** | **%** |
| Docker | 1076 | 11.95% | 27924 | 7.64% |
| Kubernetes | 771 | 8.56% | 10084 | 2.76% |
| Terraform | 692 | 7.69% | 5412 | 1.48% |
| npm | 629 | 6.99% | 28409 | 7.78% |
| Pip | 586 | 6.51% | 11778 | 3.22% |
| Homebrew | 367 | 4.08% | 11964 | 3.27% |
| Yarn | 256 | 2.84% | 13169 | 3.60% |
| Webpack | 157 | 1.74% | 13082 | 3.58% |
| Make | 480 | 5.33% | 11718 | 3.21% |
| NuGet | 213 | 2.37% | 9208 | 2.52% |

The comparison of the top 10 tools used by DevOps Specialists and General Software Developers in Table 10 highlights distinct usage patterns between the two groups. Docker is notably more popular among DevOps Specialists, with 1,076 users (11.95%) compared to its usage by 27,924 General Developers (7.64%), emphasizing its vital role in DevOps practices. Similarly, Kubernetes is employed by 771 DevOps Specialists (8.56%), significantly more than its usage by 10,084 General Developers (2.76%), emphasizing its importance in container orchestration, a key element of DevOps. Terraform, which facilitates building, changing, and versioning infrastructure, is preferred more by DevOps Specialists (692 users, 7.69%) than General Developers (5,412 users, 1.48%), aligning with the DevOps emphasis on infrastructure management. These differences in tool usage patterns reflect the specific requirements and practices of the DevOps field. Conversely, tools like npm, Pip, and Homebrew show a more balanced distribution, indicating their broader application across different software development practices.

The data in Table 11 outlines the prevalent DevOps-related practices among respondents, categorized into DevOps Specialists and General Developers. Continuous Integration (CI) and Continuous Delivery (CD) practices are widely adopted, with 20% usage reported by both groups. Automated Testing shows near equal usage, with 16% of DevOps Specialists and 17% of General Developers employing this practice. Microservices adoption is 14% for both groups, indicating its significant role in modern software development. The usage of Observability Tools and Developer Portals exhibits slight variations, with higher usage among DevOps Specialists. The Innersource Initiative sees an equal, modest adoption rate of 4% in both groups, reflecting interest in collaborative development practices. AI-Assisted Technology Tools are slightly more prevalent among General Developers (4%) than DevOps Specialists (3%). A small subset of respondents indicated not using any listed practices. These findings underscore commonalities in practices such as CI/CD, automated Testing, and microservices while highlighting variations in adopting observability tools and DevOps functions, emphasizing distinct focuses and priorities within their respective work practices.

**Table 11:** Top DevOps-related practices *Among Survey Respondents*

| Technology | DevOps Specialists Count | DevOps Specialists % | General Developers Count | General Developers % |
|---|---|---|---|---|
| Continuous Integration (CI) and Continuous Delivery | 763 | 20% | 29,293 | 20% |
| DevOps Function | 801 | 20% | 24,457 | 17% |
| Automated Testing | 615 | 16% | 24,783 | 17% |
| Microservices | 550 | 14% | 19,976 | 14% |
| Observability Tools | 532 | 14% | 15,910 | 11% |
| Developer Portal or Other Central Places to Find Tools/Services | 339 | 9% | 14,996 | 10% |
| Innersource Initiative | 167 | 4% | 5,739 | 4% |
| AI-Assisted Technology Tool(s) | 126 | 3% | 6,417 | 4% |
| None of These | 17 | 0% | 4,967 | 3% |

## 5. Discussion

The 2023 Stack Overflow Survey analysis elucidates several pivotal distinctions and parallels between DevOps Specialists and General Software Developers. Both cohorts exhibit a youthful demographic, predominantly in the 25-34 age range, indicative of a younger workforce in the technology industry. However, there is also a significant presence of individuals aged 35-44 among both groups, suggesting that this field benefits from older and more experienced professionals. The United States and India emerge as significant contributors in both sectors, reinforcing their pivotal roles in the global tech landscape. A notable trend in the educational backgrounds of these professionals is the emphasis on higher education, particularly bachelor's and master's degrees. This trend underscores the industry's requirement for academically trained professionals.

The professional attributes examined – coding experience, employment status, and remote work preferences – reveal a varied yet substantial experience across both categories. Many professionals have accrued considerable years in the field, predominantly full-time, suggesting a stable and committed career trajectory in these areas. The shift in work modes post-pandemic is conspicuous, with many professionals in both categories adopting remote or hybrid models. This evolution indicates a move towards more flexible work environments in the tech industry, potentially leading to greater geographic diversity and varied working styles. These models also affect collaboration dynamics, necessitating innovative strategies for effective team management and communication.

Understanding these demographic trends is crucial for organizations in recruitment, team composition, and professional development planning. For instance, the younger demographic among General Developers suggests a need for continuous learning opportunities and mentorship programs. Conversely, the mid-career status of many DevOps professionals may necessitate advanced training and leadership development.

The transition towards remote and hybrid work environments calls for reevaluating work practices to sustain productivity and effective collaboration in distributed settings. Investments in tools and technologies that facilitate remote work and adapted management practices for geographically dispersed teams are becoming increasingly important.

This demographic analysis underscores the need to consider age, education, and work preferences in shaping the software development sector's policies, training programs, and work environments. By understanding these demographic elements, industry leaders, educators, and policymakers can more effectively cater to the evolving needs of the software development community.

The educational backgrounds of DevOps Specialists and General Software Developers show no significant differences, indicating a broad spectrum of educational qualifications in both groups. A substantial number of DevOps Specialists and General Developers hold bachelor’s and master’s degrees, which reflects a similar level of specialized knowledge across both domains. This includes expertise in areas like infrastructure, automation, software engineering, and programming. Additionally, both groups include individuals with a range of educational experiences, from some college or university study to associate degrees and secondary school education. This diversity suggests that there are multiple pathways into the software development profession, regardless of whether one is a DevOps Specialist or a General Developer.

The survey's findings on the preferences and practices of DevOps specialists and general software developers are telling. DevOps professionals favor tools like Docker, Kubernetes, Terraform, and Ansible, underscoring their focus on optimizing operational workflows. On the other hand, General Developers demonstrate a broader spectrum of tool usage, from Docker and npm to Pip and Homebrew, addressing diverse development scenarios.

The analysis of tool usage and technical skills among DevOps professionals and General Software Developers indicates a convergence in their expertise, highlighting the complementary and interdependent nature of their roles within the software development ecosystem. Both groups exhibit proficiency in a diverse range of areas, including but not limited to automation, CI/CD, infrastructure management, and various programming and development skills. This convergence reflects the holistic and continuously evolving nature of roles in software development, emphasizing the necessity for a blend of specialized expertise and broad knowledge across different positions in the field. Furthermore, the lack of significant divergence in tool and technology preferences between these groups reinforces their interdependent roles, suggesting that both categories of professionals are integral to the collaborative dynamics of the industry.

The survey results reflect broader industry trends, with DevOps emerging as a distinct field in response to the need for efficient, rapid, and continuous software delivery. These findings reinforce the industry's shift towards integrating development and operational processes, emphasizing automated, reliable, and scalable software development methods.

For educators, industry professionals, and policymakers, these insights are invaluable. They highlight the need to align educational curricula with industry demands, emphasize the importance of team building and resource allocation in software projects, and stress the necessity for continuous skill development, particularly in automation and cloud technologies. These insights are crucial for competitiveness in the rapidly evolving software development landscape.

## 6. Conclusion

The study offers comprehensive insights into the evolving landscape of software development, mainly focusing on the roles of DevOps specialists and general software developers. It is revealed that there is no notable divergence in the preferences for tools and technologies between these two groups, indicating their interdependent and complementary roles in the industry. It delves into demographic trends within these professional groups. Generally, software developers tend to be younger, which might suggest greater adaptability or inclination toward emerging technologies and methodologies. In contrast, DevOps professionals are often found to be in the mid-career stage, implying that the intricacies and responsibilities of DevOps roles benefit from or necessitate a higher level of experience or specialized training.

Additionally, the increasing adoption of remote and hybrid work models by DevOps specialists and general software developers is a significant trend. This shift reflects the broader transformation in work culture across the tech industry and indicates a move towards more geographically diverse and flexible working styles. Such changes will likely influence collaboration dynamics, team management, and communication strategies, especially in distributed work environments.

DevOps specialists and general software developers demonstrate a concerted focus on enhancing automation and efficiency, as evident in their use of tools like Docker and Kubernetes. This shared orientation underscores the collaborative nature of modern software development, where diverse roles converge around standard technological solutions to optimize development processes.

These findings carry substantial implications for various stakeholders in the software development sector. For educators and trainers, the results underscore the need to tailor educational frameworks and curricula to equip upcoming developers with relevant skills in automation and cloud technologies. Industry professionals and practitioners are encouraged to remain attuned to the latest tools and technological advancements to maintain competitiveness and efficiency in their practices. Organizational leaders and policymakers should consider formulating and implementing policies that promote continuous learning and skill enhancement, particularly in areas pivotal to modern software development.

This study provides a valuable perspective on the current state and trends in software development, highlighting the importance of understanding the distinct yet complementary roles of DevOps specialists and general software developers. As the technology landscape continues to evolve, recognizing and adapting to these dynamics will be crucial for effective project management, strategic decision-making, and ensuring the sustainable growth of the software development industry.